\documentclass{optica-article}

\journal{opticajournal} 

\articletype{Research Article}
\usepackage{soul}
\usepackage{lineno}
\usepackage{graphicx}
\usepackage{dcolumn}
\usepackage{bm}
\usepackage{braket}
\usepackage[utf8]{inputenc}
\DeclareUnicodeCharacter{2212}{\textminus}
\DeclareUnicodeCharacter{0308}{\"}

\begin{document}

\title{Single-Photon Generation: Materials, Techniques, and the Rydberg Exciton Frontier}

\author{Arya Keni \authormark{1,$\dagger$}, 
Kinjol Barua \authormark{1,$\dagger$}, 
Khabat Heshami \authormark{2,3,4,*},
Alisa Javadi \authormark{5,6,*},
and Hadiseh Alaeian\authormark{1,7,*}}

\address{\authormark{1} Elmore Family School of Electrical and Computer Engineering, Purdue University, West Lafayette, IN 47907, USA \\
\authormark{2} National Research Council of Canada, 100 Sussex Drive, Ottawa, Ontario K1N 5A2, Canada \\
\authormark{3} 
Department of Physics, University of Ottawa, ON K1N 6N5, Canada\\
\authormark{4} Institute for Quantum Science and Technology, Department of Physics and Astronomy, University of Calgary, Alberta T2N 1N4, Canada\\
\authormark{5} Department of Electrical and Computer Engineering, University of Oklahoma, Norman, OK 73069, USA \\ 
\authormark{6} Department of Physics, University of Oklahoma, Norman, OK 73069, USA \\
\authormark{7} Department of Physics and Astronomy, Purdue University, West Lafayette, IN 47907, USA \\
}

\authormark{$\dagger$ These authors contributed equally.}\\
\email{\authormark{*} khabat.heshami@uottawa.ca, alisa.javadi@ou.edu, halaeian@purdue.edu}

\begin{abstract*} 

Due to their quantum nature, single-photon emitters generate individual photons in bursts or streams. They are paramount in emerging quantum technologies such as quantum key distribution, quantum repeaters, and measurement-based quantum computing. Many such systems have been reported in the last three decades, from Rubidium atoms coupled to cavities to semiconductor quantum dots and color centers implanted in waveguides.  This review article highlights different material systems with deterministic and controlled single photon generation. We discuss and compare the performance metrics, such as purity and indistinguishability, for these sources and evaluate their potential for different applications. Finally, a new potential single-photon source, based on the Rydberg exciton in solid−state metal oxide thin films, is introduced, briefly discussing its promising qualities and advantages in fabricating quantum chips for quantum photonic applications.

\end{abstract*}


\section{Introduction}
\subsection{Applications of single-photon sources}~\label{sec:intro}
Single photons are flying qubits with crucial roles in applications such as all-optical quantum computing~\cite{Browne2005, OBrian2007}, long-distance quantum communication~\cite{sangouard2011, azuma2023}, and cryptography~\cite{pirandola2020}. They are generally classified into two categories: probabilistic and on-demand. Probabilistic sources, such as those based on spontaneous parametric down-conversion (SPDC), have been the most widely used to date~\cite{Kaneda2016, Guo2016,  Tiedau2019, Lange2022}. Although these sources have enabled significant achievements, including some of the first tests of Bell inequalities~\cite{Tittel1998, Tittel1998-2, Tsujimoto2018}, their probabilistic nature introduces a considerable overhead for large-scale experiments. Moreover, there is a negative trade-off between the purity of the source and the probability of photon generation which limits the source efficiency to only a few percent.

On-demand SPEs, on the other hand, rely on the spontaneous emission from quantum emitters and can achieve near-unity photon creation efficiency, making them attractive for scalable applications. So far, most on-demand single-photon sources involve a single or ensemble of emitters coupled to resonant cavities or waveguides, and creating a deterministic single-photon source without relying on resonant optical structures remains a major challenge. On the other hand, there are cavity-free protocols, such as the Duan-Lukin-Cirac-Zoller (DLCZ), which use heralded preparation of collective spin excitations in atomic ensembles, later converted into photons via stimulated Raman processes~\cite{Duan2001}.

Deterministic sources based on solid-state quantum emitters are particularly attractive as they generally benefit from faster generation rates and ease of operation compared to their atomic counterparts. In solid-state systems, atom-like sources can be excited using pulsed lasers to emit single photons at GHz rates sequentially. Furthermore, spontaneous emission can be accelerated via nanophotonics to exceed 10 GHz, which also allows directing the photons to well-defined optical modes of cavities and fibers (cf. Sec.~\ref{sec: existing SPE} for more information). These sources are being developed from a variety of systems including quantum dots, color centers in diamonds, nanostructures like carbon nanotubes~\cite{Hofmann2013}, O$_{2}$ vacancies in Oxides such as Zinc Oxide (ZnO)~\cite{Chen2019}, as well as local defects and excitons in transition metal dichalcogenides~\cite{Schwarz2016, Parto2021, Abramov2023, Piccinini2024}, and two-dimensional materials~\cite{Dietrich2017-2, Akbari2021, Cassabois2022, Smit2023}.

Besides, solid-state quantum emitters offer flexibilities in terms of the emission wavelength and the generation of quantum states of photons. Different emitters have unique energy levels that affect emission dynamics, influencing photon quality and applicability, enabling the deterministic generation of entangled photonic states essential for quantum computing, and quantum repeaters. Some emitters even have spin states that can be monitored at room temperature, enabling applications such as nanoscale magnetometry for biological systems. The emergence of new emitters could transform the quantum technology landscape and introduce further capabilities. 

This review first examines various types of single-photon sources, with an emphasis on solid-state platforms. Then, it explores how Rydberg atoms' unique properties have been utilized within the last two decades to map single atomic excitations onto photons with well-defined spatial and temporal modes and how recently discovered Rydberg excitons in solid states hold promise for enabling a high-quality SPE similar to their atomic counterparts. The keys to translating research into real-world applications, scalability, and reproducibility, remain the primary challenges of solid-state photon sources. Large-scale, high-quality production of solid-state SPE with consistent spectral and temporal performance are current targets for the research community and a central focus for research and startups in the quantum industry. No single emitter platform meets all criteria for an ideal single-photon source across all applications. Trade-offs in performance, scalability, and application-specific requirements drive ongoing exploration of new materials and platforms.

\subsection{Organization of the review}
The structure of this review is organized as follows: in Sec.~\ref{sec: characterization}, we begin with a comprehensive overview of single-photon emitter metrics, including aspects of efficiency, photon purity, and indistinguishability. Section~\ref{sec: existing SPE} examines commonly used SPEs, ranging from single atoms to excitons, which primarily rely on single-particle effects. This section also explores the integration of solid-state sources with nanophotonic cavities and waveguides to enhance SPE characteristics. In Sec.~\ref{sec: Rydberg}, we introduce a distinct category of single-photon sources based on a many-body effect, known as the Rydberg blockade. We provide a brief overview of Rydberg atoms, emphasizing the impact of strong, long-range dipole-dipole interactions that facilitate deterministic optical nonlinearity at the single-photon level, and discuss how this feature can support the creation of on-demand, efficient SPEs. Additionally, we present recent advances in Rydberg systems centered on excitons in solid-state systems. By discussing the properties of Rydberg excitons and proposing methods for generating SPEs, we highlight the potential of this platform in advancing solid-state quantum optics and technologies. Lastly, in Sec.~\ref{sec: conclusion}, we conclude the review and highlight several active research directions in single-photon source development.

\section{Characterization of single$-$photon sources}~\label{sec: characterization}

\begin{figure}
    \centering
    \includegraphics[width=\textwidth]{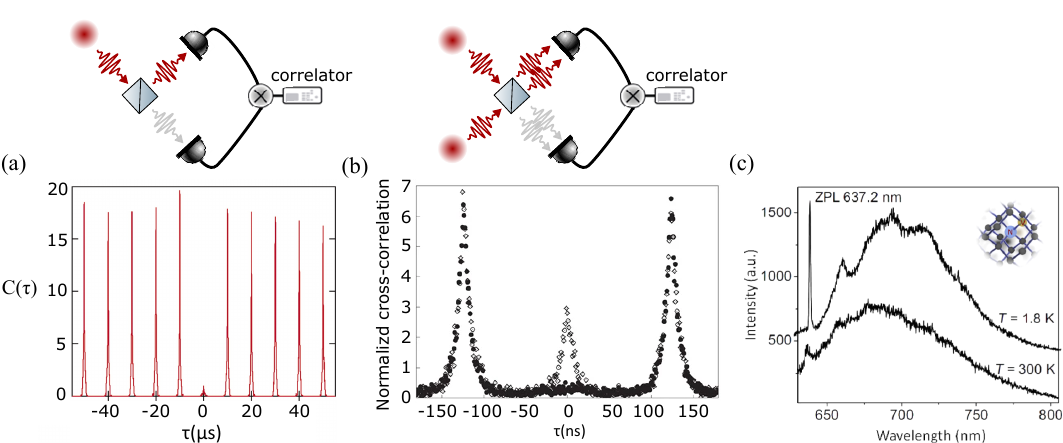}
    \caption{\label{fig: HBT} Various measures of single-photon sources. (a) Autocorrelation function $g^2(\tau)$ of light emitted from a single atom. The photon stream is split into two paths, and the photon arrival times at the detectors are correlated. The absence of simultaneous detection events at $\tau=0$ indicates an antibunched photon source. (b) Indistinguishability measurement of photons from two ions. Photons are launched into the two input ports of the beamsplitter. Indistinguishable photons arriving at the beamsplitter simultaneously follow the same path, resulting in a dip in the correlation between the two detectors at $\tau=0$ (black circles). In contrast, distinguishable photons lead to correlated detection events at $\tau=0$ (light rectangles). (c) The emission spectrum of a negatively charged NV center at two different temperatures. The sharp peak around 637 nm corresponds to the zero$-$phonon line emission of the NV center, while the broad emission on the lower wavelength side is attributed to phonon$-$assisted emission events. {Panel (a) adapted from reference figure \cite{Mckeever2004}, (b) from figure \cite{Maunz2007}, and (c) from figure \cite{Beha2012}. }} 
\end{figure}

Deterministic sources discussed in this review generally rely on spontaneous emission from quantum emitters having a mono-exponential form ($I(t)\propto e^{-\gamma t}$), where $\gamma$ is the decay rate of the emitter. $\gamma$ widely varies for different emitters, ranging from hundreds of kHz for cold atoms to several gigahertz for solid-state emitters. High decay rates are generally desirable as the emitter's decay rate directly influences the photon generation rate. Furthermore, faster optical emission rates help overcome adverse processes that cause dephasing or nonradiative decay channels.

An ideal single-photon stream must be free of multi-photon states. The key metric used to quantify the \emph{purity} of an SPE is the equal-time second-order correlation function,  $g^{(2)}(0)$, which quantifies the absence of multi-photon events in a stream of photons. Hanbury-Brown-Twiss (HBT) measurements are commonly used to characterize the purity of SPEs. As depicted in Fig.~\ref{fig: HBT}(a), the setup includes a 50-50 beamsplitter followed by two single-photon detectors. While $g^{(2)}(0)$ values below 0.5 indicate single-photon emission, an ideal source should have $g^{(2)}(0)$ close to zero; $g^{(2)}(0)$ close to $1\%$ can be achieved with most quantum emitters, in the best case, $g^{(2)}(0)$ can be as low as $10^{-5}$. $g^{(2)}(0)$ values slightly greater than zero are observed due to imperfections, such as background noise, laser leakage, and multi-photon emission events~\cite{Fischer2017, Das2019}. When investigating the purity of an SPE, it is crucial to measure $g^{(2)}(0)$ with excitation powers close to saturation for continuous wave excitation and close to the $\pi-$pulse excitation, i.e. full inversion, for pulsed measurements so that the measurements closely reflect the performance of a deterministic SPE.

Another important feature of single-photon sources is their ability to generate \emph{identical} photons or \emph{indistinguishable} photons. Indistinguishability is typically quantified through the Hong-Ou-Mandel (HOM) effect; two identical photons impinging on a beam splitter will exit the beam splitter through the same arm. Fig.~\ref{fig: HBT}(b) shows the configuration of a HOM setup as well as the measurement outcomes for the case of identical photons (filed markers) and photons that have orthogonal polarization (empty symbols). The contrast between the two measurements reveals the indistinguishibility of generated photons. The interference visibility is defined as the ratio of coincidence events around zero time delay ($M=1-\frac{A_{\|}}{A_{\perp}}$), where $A_{\|}$ is the number of coincidence events when the photons entering the beamsplitter have identical polarizations, and $A_{\perp}$ is the number of events when the two streams have perpendicular polarization. This metric is particularly valuable in photonic quantum information processing as the HOM effect is the operating principle for photonic quantum gates. HOM visibility values typically range from 0 to 1, with 0 signifying highly distinguishable photon properties (such as polarization or spectral), and 1 indicating completely indistinguishable photons. For SPEs, typical HOM visibility values range from 0.60~\cite{Sipahigil2012, Kaplan2023, Chen2022} to 0.995~\cite{Somaschi2016, Uppu2020, Zhai2022}.

Measuring HOM interference visibility in the same way as depicted in  Fig.~\ref{fig: HBT}(b) is challenging as both emitters have to be identical and stay identical throughout the measurement, making the system highly sensitive to the parameter in the environment of the two emitters. It is only feasible with highly pure and stable quantum emitters such as atoms and ions or a limited number of solid-state ones~\cite{Zhai2022}. However, HOM interference visibility between photons from the same source can also be measured using a Mach-Zehnder interferometer. This method removes the sensitivity to slowly varying factors, such as column noise arising from fluctuating charges around quantum emitters.

The performance of solid-state quantum emitters is further complicated due to their stronger interaction with the environment. Two further parameters characterize these emitters: quantum efficiency (QE) and the Debye-Waller factor (DBW). QE is the percentage of excitation events that result in the emission of photons from the emitter. QE changes widely for different emitters, from near unity values for quantum dots at cryogenic temperatures to a few percent for color centers at room temperature. For most solid-state quantum emitters, QE is significantly lower at room temperature, often around 50-70\% even in the most optimal conditions due to increased phonon-electron scattering, spectral diffusion, and thermal effects~\cite{binnemans2015, Castelletto2013}. 

In the context of phonon-induced dephasing, a key parameter to consider is the comparison between light emitted into the zero-phonon line (ZPL) and the phonon side-band (PSB), as depicted in Fig.~\ref{fig: HBT}(c). A common figure of merit used here is the Debye-Waller factor, which quantifies the ratio of photons emitted into the ZPL to the total photons within the entire spectrum~\cite{florian2018, Yu2021}, see Fig.~\ref{fig: HBT}(c) for a complete spectrum of a solid-state emitter including ZPL and the incoherent part of the emission. DBW factor can vary significantly across different solid-state emitters. As an example, the DBW for nitrogen-vacancy centers is around 0.03-0.05, meaning only about 3-5\% of the photons are emitted without being affected by phonons, while it reaches values close to 95\% for quantum dots at cryogenic temperature~\cite{Lodahl2015}.%

\section{Overview of existing emitters}~\label{sec: existing SPE}

\begin{figure}
    \centering
    \includegraphics[width=\textwidth]{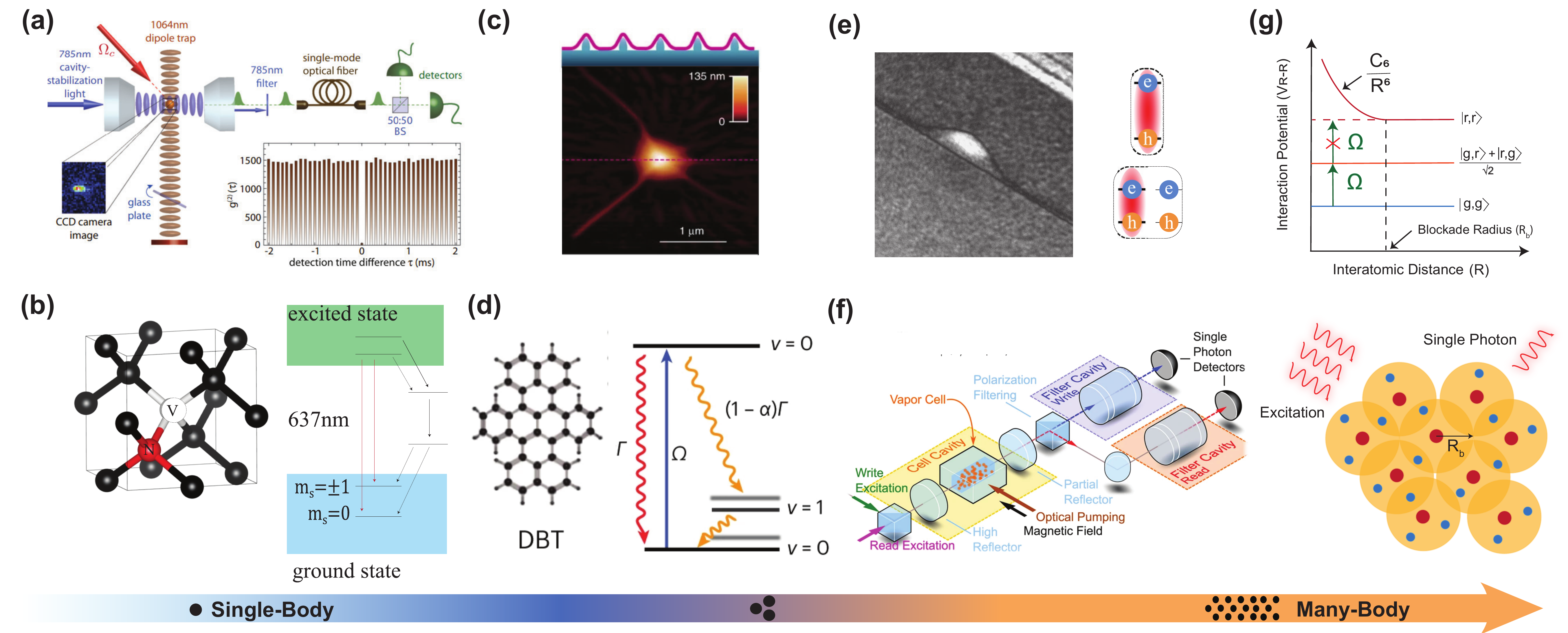}
    \caption{\label{fig: overview} A summary of on-demand single$-$photon emitters sorted according to the nature of the photon emission process from the single body emitters to many$-$body emitters. (a) A single Rubidium (Rb) atom trapped inside the high-finesse optical cavity works as a single photon source with $g^{(2)}(0)$ of 0.02~\cite{Mucke2013}. (b) Crystal structure of a nitrogen vacancy center and its level structures. The black colored spheres indicate a perfect diamond lattice, while the red sepher represents a nitrogen atom, and the missing atom is indicated with a \textit{V}. (c) Atomic force microscope image of a mono$-$layer WSe$_2$ on a templated substrate~\cite{Palacios2017}(d) Atomic structure of a dibenzoterrylene molecule along with its level structure~\cite{Lange2024}. (e) Transmission electron micrograph of an as$-$grown InAs quantum dots, along with processes involved in the generation of an exciton and a biexciton. TEM image courtesy of Jean-Michel Chauveau and Arne Ludwig.  (f) Deterministic room-temperature single-photon source with atomic vapors with anti-bunching as low as $g^{(2)}(0)$=0.2~\cite{Dideriksen2021}. (g) Top: interaction potential between two highly-excited Rydberg atoms as a function of distance (R). Bottom: Rydberg blocked in the ensemble excludes more than one Rydberg atom in a blockade radius.}  
\end{figure}

Single-photon emission can be realized with emitters with a wide range of geometrical scales. Figure~\ref{fig: overview} summarizes different SPEs according to their geometrical scales. At the fundamental limit, single-photon emission was first observed in resonance fluorescence from flying sodium atoms~\cite{Kimble1977}, which also constituted the first observation of antibunching in photon statistics. Other atomic species as well as trapped ions were later used for generating single photons~\cite{Diedrich1987}. Generally, laser-cooled atomic emitters benefit from very high optical coherence and spontaneous emission limited spectral shapes, as well as near unity quantum efficiency. The main limitation for atomic sources is the long lifetime of the excited state, typically on the order of a few microseconds for visible and near-IR transitions, which limits the photon generation rate to a few megahertz. Various aspects of single photon generation, such as collection efficiency and indistinguishability, were characterized and optimized for atomic sources. For instance, HOM measurements between photons from two trapped ions achieved indistinguishability of ~86\% (95\% when corrected for dark counts)~\cite{Maunz2007}; as shown in Fig.~\ref{fig: HBT}(b).

Defects in solid-state crystals are another class of tightly-confined SPEs. While photoemission from these centers was known in the 1960s and was used to assess the purity and quality of crystals for electronics applications, for the first time, the antibunching from these emitters was observed for a color center comprised of a substitutional nitrogen atom and an adjacent carbon vacancy (NV)~\cite{Kurtsiefer2000} in diamond; see Fig.~\ref{fig: overview}(b) for a sketch of the crystal structure of an NV center. The level structure of a negatively-charged NV center is composed of triplet ground and excited states, as depicted in Fig.~\ref{fig: overview}(b). The presence of a singlet state allows initialization and readout of the spin state of the NV center. In the last decade, antibunched photon emission from defect centers in various host materials, such as silicon carbide~\cite{Sorman2000, Falk2013}, hexagonal boron nitride~\cite{Martinez2016, Mendelson2021, Grosso2017, Tran2016}, and silicon nitride~\cite{Senichev2021, Senichev2022}, and more recently silicon~\cite{Redjem2020, Baron2022} has been observed. While NV centers are widely used in nano magnetometers~\cite{Thiel2019, Palm2024}, and have brought a transformational impact in this field, photon emission from color centers generally suffers from low QE (20\% in the case of NV center), strong interaction with vibrational states of the defect center (DBW$\approx4\%$), and the relatively slow decay rate of the excited state (around 10\, ns). Hence, single-photon sources based on color centers generally require Purcell enhancement to improve their coherence properties, as will be discussed in the following sub-section.  Furthermore, narrow spectral or temporal filtering can be employed to increase the indistinguishability of photons from color centers. However, these methods compromise the overall efficiency and brightness of the source. While $g^{(2)}(0)$ on the order of a few percent can easily be achieved with color centers, the state-of-the-art indistinguishability observed is $\approx$ 72\% for SiV centers in diamond after strong spectral and temporal filtering ~\cite{Sipahigil2014} and $\approx$ 66\% for NV centers in diamond~\cite{Sipahigil2012}.

Another class of solid-state quantum emitters is based on novel two-dimensional (2D) materials. 2D materials have attracted the attention of the research community for their exotic and highly correlated electronic states. Excitons in these materials are known to have very high recombination rates, high QEs, and strong binding energies. While single photon emission in different van der Walls structures has been observed~\cite{Srivastava2015, Chakraborty2015, He2015}, The main challenge in generating single photons from excitons in 2D materials is creating an anharmonic trap to isolate a single exciton. A local defect or a strain can be used to isolate these excitons at a slightly different energy than the bulk excitons~\cite{Kumar2015}. A popular approach is based on pre-patterning a substrate and placing the exfoliated layers of TMDs on top of the pattern~\cite{Branny2017, Palacios2017}, or using an AFM tip to create a strain traps~\cite{Rosenberger2019}, see Fig.~\ref{fig: overview}(c) for a sketch as well as an AFM image of the resulting structures. Typically, the lifetime of these emitters is about 1~ns and $g^2(0)$ around 2\% have been achieved with these SPEs; however, the coherence and indistinguishability of the photons emitted from these excitons are currently limited by dephasing and charge noise. Another interesting aspect of the 2D-based single-photon sources is the availability of a wide range of materials spanning broad wavelengths. While the majority of the works so far demonstrated SPEs in visible and near-IR regions, single-photon emission from these excitons at the telecom band was recently demonstrated~\cite{Zhao2021-2}.

At a slightly larger scale, isolated dye molecules, typically composed of less than 100 atoms, are also known to produce single photons. Some of the well-studied molecules include dibenzantanthrene (DBAT) and dibenzoterrylene (DBT). These molecules are generally embedded in organic host matrices to protect them from photobleaching as well as environmental noise~\cite{Ware1965, Toninelli2010}. These molecules generally possess simple two-level electronic structures, accompanied by additional transitions involving vibrational states of the molecule, see Fig.~\ref{fig: overview}(d) for the level structure and the atomic composition of DBT molecule. The QE of these molecules is a subject of ongoing debate and values from 35\% to 70\% are suggested for these emitters~\cite{Erker2022, Musavinezhad2023}. HOM indistinguishability measurements between photons emitted from the same molecule typically result in indistinguishability of 40-50\%~\cite{Lettow2010, Schofield2022}, while similar indistinguishability between photons generated from separate DBT molecules has been observed under pulsed excitation~\cite{Duquennoy2022}. Interestingly the resonance frequency of molecules can be tuned by applying strain~\cite{Fasoulakis2023}, electric fields~\cite{Trebbia2022}, or laser pulses~\cite{Colautti2020}, which has enabled the observation of superradiant emission from molecules~\cite{Rezai2019, Stein2020, Trebbia2022, Lange2024}.

Quantum dots (QD)s based on III-V materials are arguably the most mature solid-state platform for SPEs. Much larger than the platforms discussed so far, composed of $10^5$ atoms, quantum dots benefit from the excellent quantum efficiency typical of III-V systems. They are typically an island of small bandgap material such as indium arsenide (InAs) embedded in a matrix of larger bandgap material (e.g. gallium arsenide (GaAs), Indium Phosphide (InP)~\cite{Almeida2023}, Aluminium Nitride (AlN)~\cite{Ku2020}); Fig.~\ref{fig: overview}(e) shows a transmission electron micrograph (TEM) image of an as-grown quantum dot. Quantum dots measure around 10 nm in width and a few nanometers in height. The small size of the quantum dot leads to strong Coulomb interactions between the electron and hole forming the excitons, as well as a strong interaction between two excitons causing an anharmonic energy level diagram; the right panel of Fig.~\ref{fig: overview}(e) illustrates the energy level of a single exciton and a bi-exciton in a quantum dot. The energy difference between the emission from the bi-exciton and the single-exciton levels forms the basis for single-photon generation in quantum dots. While the first demonstrations of single-photon emission from quantum dots date back to the early 2000s~\cite{Santori2002, Michler2000, Warburton2000}, significant progress has been made on these quantum emitters in the last decade. For example, very high indistinguishability (up to 98\%) has been observed for photons generated from the same QD~\cite{Somaschi2016, Uppu2020, Tomm2021}, with collection efficiencies into a single mode fiber reaching up to 70\%~\cite{Tomm2021, Ding2023}. Another exciting development in QDs has been the demonstration of very high indistinguishability (>90\%) between separate sources without any Purcell enhancement~\cite{Zhai2022}. The main open challenge for QDs lies in their random growth position, as well as their relatively large inhomogeneous broadening on the same chip. These challenges can be addressed by pre-characterization and preselection of the quantum dots~\cite{Nowak2014, Sapienza2015, Liu2024}, as well as the use of electrical tuning or local strain tuning to tune them into resonance~\cite{Grim2019}. 

At the yet larger scale, ensembles of cold ($\rho \approx 10^8  \mu m^{-3}$) and thermal ($\rho \approx 10^{20} \mu m^{-3}$) atoms, have been utilized successfully to create efficient, on-demand SPEs; Fig.~\ref{fig: overview}(f). Although this is the densest platform of all, the inter-atomic spacing is still large enough that atom-atom interaction and many-body quantum effects are not relevant. Atomic ensemble SPEs have the same strength as individual-atom SPEs such as being identical emitters and hence creating high purity and highly indistinguishable single photons. In addition, the source brightness is enhanced by a factor of $N$, i.e. the number of atoms, and the collection efficiency is improved due to the directional emission from the ensemble. Besides, the platform offers a seamless interface with built-in atomic quantum memories for single photons~\cite{Zhao2013}. One common scheme is through DLCZ protocol in a $\Lambda$-type three-level system, which typically is comprised of two ground-state hyperfine ($\ket{g_{1,2}}$), and an excited state $\ket{e}$~\cite{Duan2001}. The protocol starts with optical pumping to shelve all atoms in $\ket{g_1}$, followed by a weak off-resonant coherent rite beam to promote one atom to $\ket{e}$. The atom can be transferred to $\ket{g_2}$ either via spontaneous decay or via another beam leading to electromagnetically-induced transparency (EIT). That way, the cloud becomes projected into a coherent spin state (limited by the coherence of the ground-state hyperfine levels), storing one excitation. On-demand, the spin excitation can be transferred into a single photon by time-reversing the process. Employing this technique within the last two decades, SPEs in cold Cs cloud with a purity of 0.24~\cite{Chu2004}, cold Rb cloud with $g^{(2)}(0) \approx 0.14$~\cite{Chanelire2005, Chen2006}, thermal Rb vapor with $g^{(2)}(0) \approx 0.3$~\cite{Eisaman2005, Dideriksen2021}, have been reported. Four-wave mixing (FWM) is another common scheme extensively used in both cold and thermal atomic vapors to create single photons~\cite{Fan2009, Liu2017-Sci, Gu2019, Caltzidis2021}.

\section{single-photons based on Rydberg super-atoms}~\label{sec: Rydberg}

\begin{figure}
    \centering
    \includegraphics[width=\textwidth]{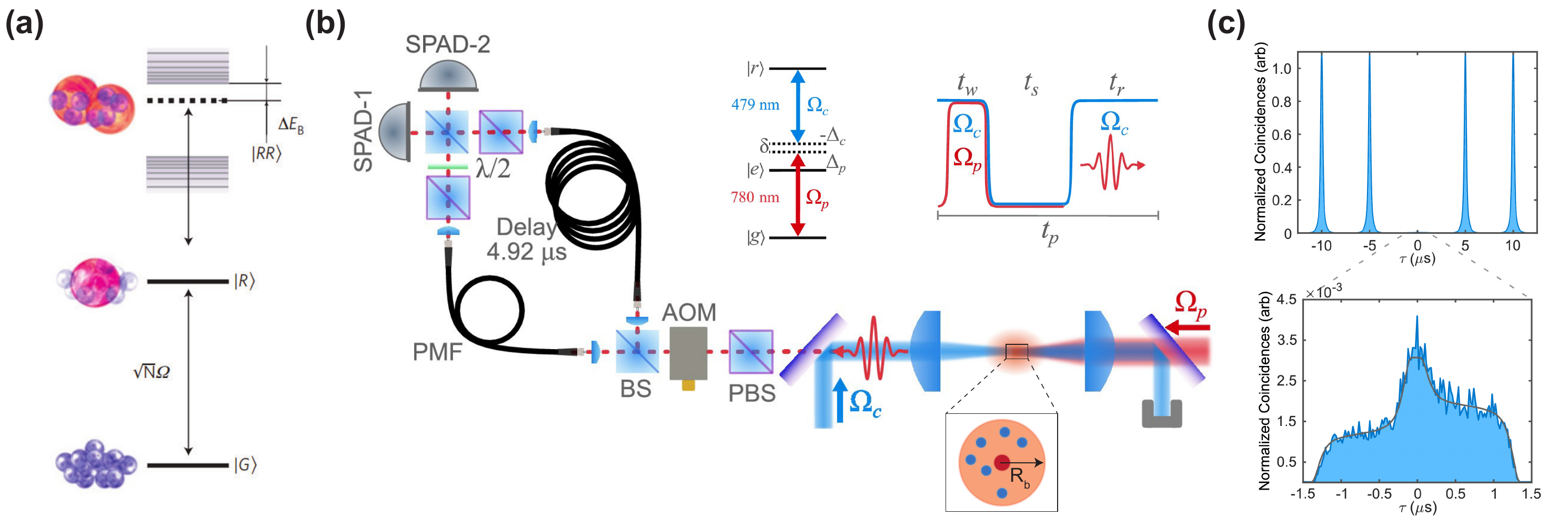}
    \caption{\label{fig: Rydberg} Single-photon source based on Rydberg super-atom. (a) Level diagram and laser excitation scheme for creating Rydberg atoms. (b) The experimental setup with two excitation beams and the Rydberg cloud creates single photons following the pulsing sequence in the top-right corner. The output light passes through a polarizing beamsplitter (PBS) to separate the emitted photons based on the polarization followed by an acousto-optic modulator (AOM) to suppress photons during the write period. The polarization and temporally-filtered photons were sent to a beamsplitter (BS) followed by two polarization-maintaining fibers (PMF) for further spatial mode filtering before photon purity measurements through HBT. (c) The top panel shows the $g^{(2)}(0)$ measurements clearly showing the suppression of the two-photon events at zero delays. The lower panel shows the zoomed-in auto-correlation signal close to zero highlighting the high purity of the SPE. Schematics and experimental results are from~\cite{Ornelas2023}.}  
\end{figure}
At the mesoscopic scale, Rydberg atoms are the largest quantum particles with unique exotic features such as interaction ranges exceeding tens of $\mu$m, as depicted in Fig.~\ref{fig: overview}(g). Rydberg atoms have valence electrons that are highly excited and far from the nucleus, with a large principal quantum number $n$. Compared to ground-state atoms, Rydberg atoms exhibit amplified characteristics, such as huge electric dipole moments and long-range interactions. The most striking features of Rydberg atoms stem from the spatial extent of electron wavefunction. The average spatial footprint of the wavefunction scales as $n^2$; $n$ being the principal quantum number associated with the state. The first direct consequence of this property is that the wavefunction overlap between the ground state and a high-lying Rydberg state gets increasingly small with higher $n$. This leads to longer lifetimes (narrower absorption lines) for Rydberg states typically scaled as $n^3$. On the other hand, the overlap between Rydberg states with the same or similar principal quantum number becomes increasingly large, resulting in the transition dipole moment to scale as $n^2$. The increasing transition dipole moment of Rydberg states enables Rydberg-Rydberg interaction through dipole-dipole interaction $\propto$ $n^4/R^3$ and another process known as the van der Waals interaction scaling as $n^{11}/R^6$~\cite{Sibalic2018}.

These large dipole moments allow significant interactions with weak electromagnetic fields. Using static electric or magnetic fields, lasers, or microwaves, these interactions can be finely tuned, making Rydberg atoms particularly effective for creating a controllable testbed to study the quantum many-body phenomena. The strength of interactions between atoms can vary significantly depending on the principal quantum number $n$ of the atomic state. As depicted in the top panel of Fig.~\ref{fig: overview}(g), when one atom is excited to the Rydberg state, it shifts the energy levels of all nearby atoms off-resonance. This effect, known as the \emph{Rydberg blockade}, creates conditional dynamics that are highly desirable for quantum information processing. As a result, if two atoms are located within the blockade radius $R_b$, they cannot both be simultaneously excited to Rydberg states.

In a volume with radius $R_b$ containing $N$ atoms, the singly-excited state emerges where all the atoms within this volume collectively share a single atomic excitation (see the lower panel of Fig.~\ref{fig: overview}(g)), formally described as 

\begin{equation}
    \ket{W} = \frac{1}{\sqrt{N}}\sum_{k = 1}^N \ket{g_1 g_2 \cdots r_k \cdots g_N}\, ,
\end{equation}
where $k$ indicates the excited atom in the Rydberg state. 

This highly-entangled state is typically called a \emph{superatom}. In other words, an ensemble of atoms in the blockaded volume can be viewed as an effective two-level Rydberg superatom. In an ordinary nonlinear medium, the optical response scales
linearly with the density as atoms are independent. Due to dipolar interactions, however, the optical response becomes proportional to the number of atoms in a blockade volume, leading to a nonlinear scaling with density. Utilizing this unique feature, there have been several theoretical proposals and experimental demonstrations of on-demand, high-purity SPEs in Rydberg ensembles~\cite{Petrosyan2018}.
Recent experiments have demonstrated the generation of a many-body excitation with no more than one Rydberg atom within a mesoscopic ensemble of ultracold atoms when the principal quantum number exceeds 70~\cite{Dudin2012}.
In~\cite{Ornelas2023}, and as depicted in Fig.~\ref{fig: Rydberg}, the authors have used a cold atomic cloud of $^{87}$Rb in an optical dipole trap followed by a two$-$photon transition to excite the atom to the Rydberg state with $n = 139$ (cf. Fig.~\ref{fig: Rydberg}(b) for detailed experimental scheme showing the excitation beams and their temporal sequences). At this highly$-$excited level, $R_b \approx 60~\mu$m is larger than the cloud size both in the plane and in the direction of propagation. At the end of the storage time, the Rydberg exciton can be emitted as a single photon via a coherent pulse, as shown in Fig.~\ref{fig: Rydberg}(c) top panel. Using this scheme the authors demonstrated a single$-$photon source with $g^{(2)} = 5 \times 10^{-4}$ as depicted in the zoomed-in plot in the bottom panel of Fig.~\ref{fig: Rydberg}(c), same$-$source photon distinguishability of $0.980$, and an average photon rate of $1.18 \times 10^4 s^{-1}$.

\subsection{Solid-state Rydberg excitons}~\label{sec: Rydberg Exciton}

\begin{figure}
    \centering
    \includegraphics[width=\textwidth]{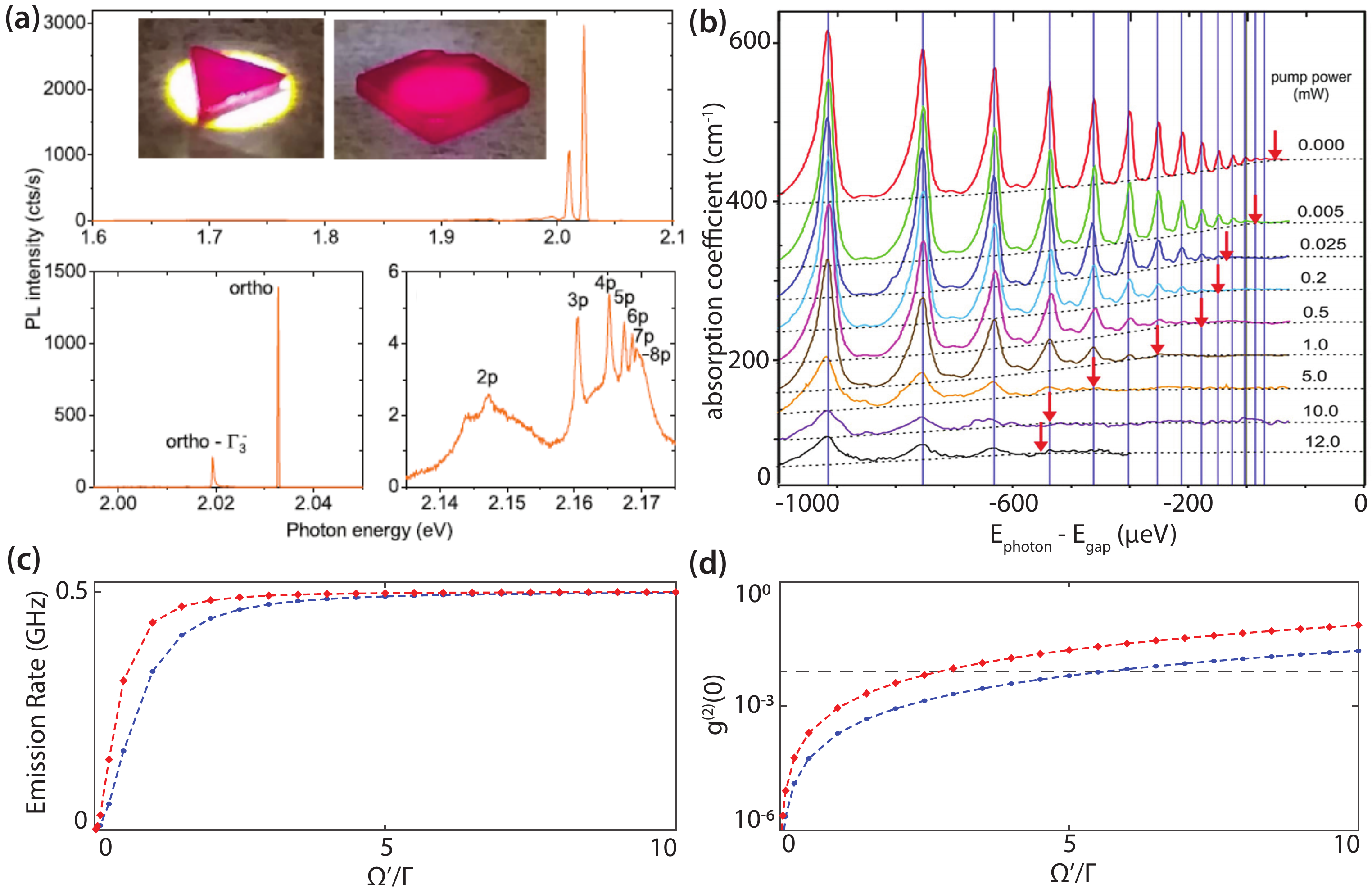}
    \caption{\label{Fig: Cu2O} Rydberg Excitons in solid-state crystal. (a) The top panel shows a polished Cu$_2$O slab and its relevant photoluminescence signal in the phonon replica (bottom left) and Rydberg region (bottom right)~\cite{Versteegh2021}. (b) The absorption coefficient of Rydberg excitons in Cu$_2$O slab as a function of pump power (exciton density) indicates the interaction effect~\cite{Kazimierczuk2014}. (c) Emission rate and (d) purity of the emitted single photon, as a function of driving Rabi frequency $\Omega'$ for a Cu$_2$O slab with the thickness of $L = 4~\mu$m (blue circles) and $L = 6~\mu$m (red diamonds) for the quantum principal number n = 24~\cite{Khazali2017}. } 	
\end{figure}

Bound states of electron-hole pairs in semiconductors exhibit similar qualities as hydrogen-like atoms. In particular, excitons in semiconductors with relatively high binding energy have high-lying bound states governed by the physics of Rydberg states~\cite{Sibalic2018, Gallagher1988}. 
The semiconductor cuprous oxide (Cu$_2$O) is well known for its relatively high binding energy of $\approx 100~meV$ and as a candidate for observation of Rydberg excitons~\cite{gross1956, matsumoto1996}. Despite decades-long interest in Cu$_2$O and its electro-optical properties, the observation of high-lying Rydberg states and Rydberg-induced non-linearity remained unexplored until 2014 when Kazimierczuk et al. observed Rydberg state of excitons up 25P in Cu$_2$O through resonant laser absorption spectroscopy~\cite{Kazimierczuk2014}. Figure~\ref{Fig: Cu2O}(a) shows a slab of Cu$_2$O (top panel) with the relevant non-resonant photoluminescence signal with the phonon replica range (left bottom panel), and the Rydberg exciton range (right bottom panel) highlighting levels up to 8P~\cite{Versteegh2021}. The dependence of the observed optical density on the laser power, as shown in Fig.~\ref{Fig: Cu2O}(b), indicated a strong dipole-dipole interaction that points towards the Rydberg blockade effect where the presence of an excited Rydberg exciton (atom) can prevent a secondary excitation by moving other Rydberg excitons (atoms) out of resonance for the probing laser~\cite{Kazimierczuk2014}. 

Such a strong non-linearity motivates the application of Rydberg excitons in quantum and non-linear optics. The advances in Rydberg atoms have already led to the development of Rydberg-based RF sensors~\cite{meyer2020}, quantum simulators~\cite{weimer2010, bernien2017}, and single-photon sources~\cite{Ornelas2023}. Rydberg excitons in Cu$_2$O demonstrate remarkable properties~\cite{assmann2020} that could potentially enable solid-state sensors~\cite{pritchett2024,heckotter2024}, quantum non-linear optics~\cite{Walther2018}, quantum simulators~\cite{poddubny2020, taylor2022}, and single photon sources~\cite{Khazali2017}. Synthesizing high-quality samples and electro-optical microstructures will be essential for demonstrating these practical applications. Examining the feasibility of these potential applications is often the first step.

In~\cite{Khazali2017}, the authors explored the effectiveness of the Rydberg blockade effect in copper oxide to generate single photons. This includes estimating the single and double exciton excitation probabilities under a range of pump intensities and other parameters such as crystal size, principal quantum number, and detuning. In terms of emission rate (brightness of such sources) and purity, the results (depicted in Fig.~\ref{Fig: Cu2O}) are promising and motivate the device, particularly the Rydberg blockade effect, derived from continuous wave driving of an effective two-level system. The steady-state solution to the exciton population is given by 
\begin{equation} \label{SingleExcProb}
\rho_{ee}(t\rightarrow \inf) =\frac{\Omega'/4}{\Delta\omega^2 + (\Gamma^2/4 + \Omega'^2/2)}\, ,
\end{equation}
where $\Delta\omega$ is detuning between the pump laser and the resonant frequency of the exciton, $\Omega'\sqrt{N}\Omega$ is the effective Rabi frequency, and $\Gamma$ is the spontaneous emission rate. Addressing and isolating a single Rydberg exciton may be challenging, and the collective enhancement that is expected from an ensemble of excitons may be eliminated. On the other hand, one should account for the probability of exciting more than one exciton will affect the purity of the emission from such a source. In fact, in the absence of exciton-exciton interaction (nonlinearity in the optical response) the emission statistics will be the same as the driving field (Poissonian distribution). The Rydberg-Rydberg interaction and in particular the Rydberg blockade effect are necessary to create a high-purity single photon source from such a system. To estimate the purity of such a source, one requires to calculate the probability of doubly excited states, $|2\rangle = \sum_{ij}|g_1 \cdots r_i \cdots r_j \cdots g_N\rangle$, in addition to the collective single excitation probability for $|1\rangle=\sum_{i} |g_1 … r_i … g_N\rangle $ which is given by Eq.~\ref{SingleExcProb}. 

The steady-state solution to the dynamics of two driven and interacting Rydberg excitons gives
\begin{equation} \label{DoubleExcProb}
\rho_{2_{ij}2_{ij}(t\rightarrow \inf)} =\frac{N\Omega^2/\Gamma^2}{1 +2 N\Omega^2/\Gamma^2} \frac{Y}{1+2Y}\, , 
\end{equation}
where $N$ is the effective number of excitons that are addressable, $Y=\frac{\Omega^2/N}{V_{ij}^2 + \Gamma^2/4}$, and $V_{ij} = \frac{C_3}{R_{ij}^3}$ is the dipole-dipole interaction with between two Rydberg excitons at a distance of $R_{ij}$. The overall double excitation probability is then given by $P_{rr}=\sum_{i<j}\rho_{2_{ij}2_{ij}}$. This allows us to evaluate expected purity as $g^{(2)}(0)= \frac{P_{rr}}{2\rho_{ee}}$ which is demonstrated in Fig.~\ref{Fig: Cu2O}(d). These preliminary evaluations motivate further development of microstructures~\cite{Steinhauer2020, Delange2023, Paul2024} and other semiconductors with high binding energy~\cite{jadczak2019, hill2015} for novel photon sources with the potential for high rates, high purity, and photonic integration for scalability.

\section{Outlook}~\label{sec: conclusion}

\begin{figure}
    \centering
    \includegraphics[scale=0.45]{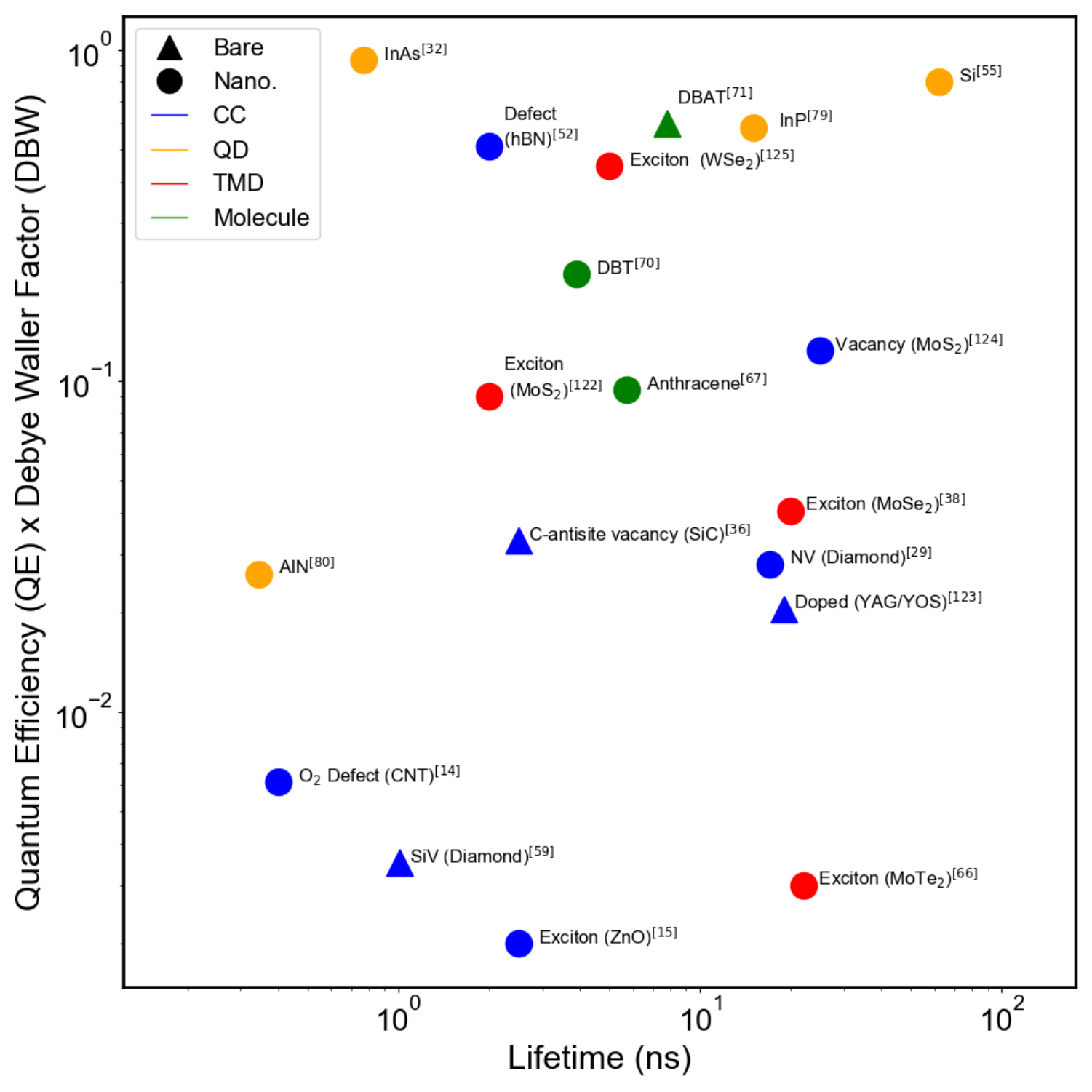}
    \caption{\label{fig: comparsion} Solid State SPE. Color center (CC), molecules, Quantum Dot (QD), and Transition Metal Dichalcogenide (TMD) are highlighted. The comparison is based on the QE, DBW, and Lifetime metric for a heuristic comparison based on applications, limitations, and material properties from a quantum optical perspective. An important consideration of nanophotonic coupling is also depicted, which causes modulation of certain measured properties of the SPE. Emitters such as YAG/YOS ~\cite{Barnes2002}, MoS$_{2}$ with sulfur vacancies ~\cite{Gupta2023} and WSe$_{2}$ ~\cite{Iff2021} are also highlighted.} 	 
\end{figure}

The lifetime and quantum efficiency of the Single-photon emitters are critical parameters that define their scope and applications. Figure~\ref{fig: comparsion} compares the lifetime versus $\eta = \textrm{QE}\times\textrm{DBW}$ for different quantum emitters. As evident, shorter lifetimes on the left side of the graph are typical of more extended emitters such as excitons, while localized emitters mostly populate the right side of the graph. Generally, color centers and other highly confined emitters suffer from stronger interaction with phononic modes, which lower their DBW and, as a result, their $\eta$, while the strong confinement of the color centers enables them to operate near room temperature. 

Rydberg atoms (excitons), on the other hand, offer a novel approach, based on the blockade effect, for limiting the number of emitted photons. As this quantum many-body effect allows only one excitation in large ensembles of tens of $\mu$m$^3$, Rydberg-based SPEs often provide high-purity identical photons. While the repetition rates of Rydberg atomic systems are typically low, Rydberg excitons have a promising perspective of resolving this bottleneck. These properties would likely put Rydberg-exciton SPEs based on the top right part of Fig.~\ref{fig: comparsion}.

At present, no single-photon emitter stands out as universally ideal; each type has unique strengths and limitations. Moreover, the definition of an "ideal" source is application-dependent. For instance, in basic quantum key distribution networks, the coherence of the single-photon source is less crucial, whereas scalability and room-temperature operation are more significant, and as a result, SPEs based on color centers are well-suited for such applications. On the other hand, for applications such as fusion-based optical quantum computing where the fidelity of fusion operations relies heavily on photon indistinguishability~\cite{Bartolucci2023}, quantum dot-based SPEs might be a more favorable option.

While the discussion in this review focused on the performance metrics of bare SPEs, integrating quantum emitters in engineered photonic environments is a proven strategy to enhance their metrics. The distinct advantage of solid-state sources is their inherent compatibility and integration with nanophotonic structures. Almost all the emitters considered here have been successfully embedded in nanostructures, which has resulted in major improvements in collection efficiency and quantum efficiency. For example, coupling quantum dots to microcavity modes~\cite{Tomm2021, Somaschi2016} and photonic crystal waveguides~\cite{Liu2018, Uppu2020, Siampour2023} has enhanced collection efficiencies, with the current record coupling efficiency of 57\% for single-photon generation from a QD coupled to a microcavity mode to a single-mode fiber~\cite{Tomm2021}. Nanophotonic cavities have also been successfully used to improve the quantum efficiency and DBW for various color centers. Similarly, nanophotonic cavities have improved quantum efficiency and Debye-Waller factors for color centers. Key achievements include boosting SiV quantum efficiency to ~90\%~\cite{Zhang2018}, enhancing the NV center’s Debye-Waller factor to 40\% through Fabry-Pérot cavities{~\cite{Yurgens2024}}, and observing superradiance from SiC color centers with strong Purcell enhancement~\cite{Lukin2023}. For further details, readers are referred to focused reviews on photonic integration of various SPEs~\cite{Aharonovich2016, Lukin2020, Uppu2021, Sartison2022}. 

Despite several advantages, solid-state systems present challenges in controlling the emitter environment, such as the ultrafast dephasing caused by charge and spin noise, which is more severe for emitters close to the surface. Overcoming these obstacles requires a deeper understanding of the emitter physics and likely the development of improved cavity designs. Scalable quantum photonics with narrow linewidth resonant structures introduces further complexities. First, The need to independently tune both the emitter and the cavity doubles the complexity of parameter control. A greater challenge lies in simultaneously achieving spectral alignment among multiple SPEs and ensuring uniform decay rates. These two requirements appear to be competing requirements to a certain degree. Parametric processes based on Rydberg excitons can offer distinct advantages~\cite{pritchett2024}.

\section*{Acknowledgment}

ADK and HA acknowledge the support from the Air Force Office of Scientific Research under award number FA9550-23-1-0489.
KB and HA acknowledge the financial support from the Industry-University Cooperative Research Center Program at the US National Science Foundation under grant No. 2224960. The work was partially supported as part of the QuPIDC, an Energy Frontier Research Center, funded by the U.S. Department of Energy (DOE), Office of Science, Basic Energy
Sciences (BES), under award number DE-SC0025620
(ADK, KB, AJ, and HA). KH acknowledges research support from NSERC Discovery Grant and Alliance programs.
     
\clearpage
\newpage
\bibliography{ref.bib}
\end{document}